\documentclass[aps,prl,twocolumn]{revtex4}
\usepackage{graphicx}

\begin{document}
\title{Phenomenology of Jet Quenching in Heavy Ion Collisions}
\author{Berndt M\"uller}
\affiliation{Department of Physics, Duke University, 
         Durham, NC 27708, USA}

\begin{abstract}
We derive an analytical expression for the quenching factor
in the strong quenching limit where the $p_T$ spectrum of
hard partons is dominated by surface emission. We explore
the phenomenological consequences of different scaling laws 
for the energy loss and calculate the additional suppression 
of the away-side jet.
\end{abstract}
\maketitle

It is commonly believed that the yield for ``hard'' observables
in high-energy nuclear reactions scales as the number of binary
nucleon-nucleon (NN) collisions occurring during the encounter of
the two nuclei. This expectation applies to processes that are
characterized by a high virtuality $q^2$, for
which final state interactions are negligible, such as lepton
pair production or the total yield of heavy flavor quarks.
There is no {\em a priory} reason to expect this rule to hold
for high-$q^2$ processes, in which the final state can be
strongly modified by interactions with comovers. A well
documented example is the production of heavy vector mesons,
such as the $J/\Psi$, which is found to be ``anomalously''
suppressed in collisions of heavy nuclei at the CERN-SPS
\cite{GH99}. 

The yield of hadrons produced with high transverse momentum $p_T$
in Au+Au collisions at the Relativistic Heavy Ion Collider (RHIC)
has recently been shown to be significantly suppressed in comparison
with the cumulative yield of NN collisions \cite{PHEN02,STAR02}. 
This effect, called ``jet quenching'', was predicted to occur as a 
result of energy loss by the hard scattered partons due to interactions  
with the surrounding dense medium \cite{Bjo82,Tho91,Gyu94}. The 
theory of this energy loss has been a topic of intense research
over the past few years \cite{Zak96,BDMPS,Wie00,BDMS,BSZ00}. The
present consensus is that the dominant mechanism for the energy loss
in QCD is collisionally induced radiation of gluons by the fast parton.

It is difficult to measure the energy loss of a scattered parton 
directly in heavy ion reactions, because the large multiplicity 
of emitted hadrons makes it almost impossible to isolate 
the resulting jet by kinematic cuts. However, the energy loss of
the parton is imprinted as an equivalent loss of energy of the
leading hadron produced in its fragmentation \cite{Wan92}. This
is what has been observed in the RHIC experiments. Generally, it
is assumed that the fragmentation occurs after the parton has left
the comoving medium, and thus is described by the measured vacuum 
fragmentation functions. We will, therefore, not be concerned with
the conversion from partons to hadrons, but focus directly on the 
$p_T$-spectrum of scattered partons.

Preliminary data from Run 2 at RHIC (Au+Au at a center-of-mass energy
of 200 GeV per nucleon pair) confirm the effect observed in Run 1
and its interpretation as jet quenching \cite{QM02}. For pions with
$p_T \approx 5$ GeV/c the measured suppression factor $Q(p_T)$ is
about 1/5. There is not yet
complete agreement between different experiments about the scaling of
the hadron yields with the number of nucleons participating in the 
reaction, $N_{\rm part}$, or with the binary collision number 
$N_{\rm coll}$. The PHOBOS collaboration has presented evidence that
the yield of charged hadrons with $p_T \approx 4$ GeV/c scales like
$N_{\rm part}$ \cite{PHOB-QM}. This finding is contradicted by data 
from the PHENIX collaboration \cite{PHEN-QM1}. It is not clear
whether the discrepancy is due to different normalization methods,
different experimental acceptance, or other reasons. 

In addition, the experiments have found that the supression factors 
for mesons and baryons are quite different up to transverse momenta 
of 5 Gev/c. It was recently proposed that this phenomenon can be 
attributed to a competition between different hadron formation 
processes \cite{Fries03,Greco}, with parton recombination dominating 
at lower $p_T$ and fragmentation at higher $p_T$.  The participant 
or binary collision scaling may be complicated in the transition 
region. The situation is predicted to simplify for $p_T > 6$ GeV/c, 
where hadrons are overwhelmingly produced by fragmentation of an 
energetic parton. 

We will first show analytically that the spectrum of high-$p_T$ 
partons, in the limit of large energy loss, is dominated by partons 
emitted from the surface of the collision zone and thus scales like 
the surface rather than the volume of the interaction region \cite{Shu01}.  
We will then explore the scaling of the quenching factor $Q(p_T)$ with 
participant number and $p_T$. We finally calculate the additional 
suppression of the away-side jet and the azimuthal anisotropy of the
parton yield in noncentral collisions.

We begin by considering the loss of energy by an energetic
parton traversing a homogeneous, static medium of thickness $L$.
We assume that the geometry is given by a cylinder with radius $R$, 
as in the boost-invariant Bjorken model \cite{Bjo83} for a nuclear 
collision with impact parameter $b=0$, and that the parton moves 
in the transverse plane in the local rest frame of the medium. 
We further assume that the effective energy loss, defined as the
shift of the momentum spectrum of fast partons, depends on $p_T$ 
and $L$ in the following general way:
\begin{equation}
\Delta p_T = \eta p_T^\mu L \, ,
\label{eq01}
\end{equation}
where $\mu$ is a scaling exponent. The linear dependence on $L$ 
holds when the loss occurs in subsequent, independent interactions
with the medium. It has also been shown to be valid, when multiple
interactions are suppressed by the LPM effect, due to the steep 
fall-off of the parton spectrum with $p_T$ \cite{BDMS}. The 
traversed pathlength inside the medium for a parton created at 
transverse position ${\bf r}$ with an angle $\phi$ relative to 
the radial direction is
\begin{equation}
L(\phi) = (R^2 - r^2 \sin^2\phi)^{1/2} - r \cos\phi
\approx \frac{z^2}{2R \cos\phi} ,
\label{eq02}
\end{equation}
where $r = |{\bf r}|$, $z^2 = R^2-r^2$, and the approximation is valid
near the surface ($z\ll R$). Perturbative QCD predicts that the 
parton spectrum at moderatedly large values of $p_T$ has the form
\cite{FMS02}
\begin{equation}
\frac{dN}{d^2p_T} = N_0 \left( 1+{p_T\over p_0} \right)^{-\nu}
\label{eq03}
\end{equation} 
with a power $\nu\approx 8$ and $p_0\approx 1.75$ GeV/c.
The quenched spectrum is given by
\begin{eqnarray}
\frac{d{\tilde N}}{d^2p_T} &=& Q(p_T)\, \frac{dN}{d^2p_T} 
    \nonumber \\
&=& \frac{1}{2\pi^2R^2} \int_0^{2\pi}d\phi \int_0^R d\,^2r \,
    \frac{dN(p_T + \Delta p_T)}{d^2p_T} \, .
\label{eq04}
\end{eqnarray} 
Replacing the integration over $r$ with an integration over $z$, using 
the approximation (\ref{eq02}), and formally extending the range of 
the integration to infinity, one finds:
\begin{equation}
Q(p_T) \approx \frac{2(p_0+p_T)}{\pi R \eta (\nu-1) p_T^\mu} \, .
\label{eq05}
\end{equation}
Two things are remarkable about this result. First, the factor $R$
in the denominator reduces the scaling of the parton yield with the
size of the reaction zone by one power of $R$, from a volume to a
surface dependence.  Second, the dependence of $Q(p_T)$ on $p_T$ is
determined by the power $\mu$ governing the $p_T$-dependence of the
energy loss. For $\mu = 1$ the quenching factor $Q$ falls slowly with 
increasing $p_T$; for $\mu = 1/2$ it grows with $p_T$, implying less 
quenching at higher $p_T$.

\begin{figure}
\resizebox{0.65\linewidth}{!}{\rotatebox{-90}
          {\includegraphics{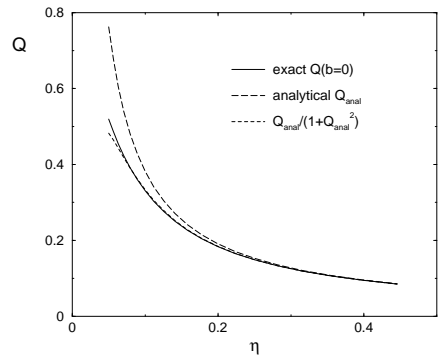}}}
\caption{Comparison between the analytical approximation (\ref{eq05})
         and the exact result (\ref{eq04}) for the quenching factor $Q$.
         The curves are for Au+Au at $b=0$ and $p_T = 6$ GeV/c, with  
         $a=1/2$. The calculation assumes a homogeneous transverse
         profile for jet production and quenching medium. Also shown
         is the expression $Q_{\rm anal}/(1+Q_{\rm anal}^2)$.}
\label{fig1}
\end{figure}

We have confirmed the range of validity of the approximate analytical
expression (\ref{eq05}) by comparing it with the exact integral (\ref{eq04}).
As seen in Fig.~\ref{fig1}, the analytical approximation $Q_{\rm anal}$
deviates from the exact result by less than 5\% when the quenching factor 
$Q \le 0.2$. An even better agreement is found when the analytical result 
is divided by the correction factor $(1+Q_{\rm anal}^2)$. The excellent 
agreement suggests that we may extend the calculation to noncentral 
collisions. Generalizing the surface-to-volume ratio of the cylindrical 
geometry ($S/V = 2/R$) to the geometry of a collision with impact 
parameter $b$, one finds
\begin{equation}
Q(p_T,b) = Q(p_T,0) \alpha_b / (\alpha_b - \sin\alpha_b)
\label{eq06}
\end{equation}
with $\alpha_b = 2 \arccos(b/2R)$. This approximation is valid with
about the same accuracy as (\ref{eq05}), except for very peripheral
collisions.

In order to be able to address the experimental data, we need to
relax some of the geometrical oversimplifications used above. 
\begin{enumerate}
\item 
The transverse profile of the primary jet yield is not homogeneous,
but proportional to the binary NN collision profile 
\begin{equation}
T({\bf r},{\bf b}) = \rho_1({\bf r}) \rho_2({\bf r}-{\bf b})
\label{eq07}
\end{equation}
where $\rho_i({\bf r})$ is the longitudinally integrated density
of nucleus $i$. 
\item
The density of the comoving medium is also not homogeneous in the 
transverse plane. We assume that it is proportional to the local 
density of participant nucleons which, in the Glauber model, is given by
\begin{eqnarray}
\rho_{\rm part}({\bf r}) 
&=& \rho_1({\bf r}) \left( 1-e^{-\sigma \rho_2({\bf r}-{\bf b})} \right)
    \nonumber \\
& & \qquad 
    + \rho_2({\bf r}-{\bf b}) \left( 1-e^{-\sigma \rho_1({\bf r})} \right) \, , 
\label{eq08}
\end{eqnarray}
where $\sigma$ denotes the inelastic NN cross section. 
\item
The comover medium expands and its density decreases with time. 
Here we assume 
\begin{equation}
\rho({\bf r},\tau) = C \rho_{\rm part}({\bf r}) / (\tau+\tau_0) \, ,
\label{eq09}
\end{equation}
which is modeled to represent a longitudinal, boost invariant expansion. 
As an approximation to the phenomenology at full RHIC energy we use the 
values $C\approx 3$ and $\tau_0 = 1$ fm/c. 
\end{enumerate}

Finally, we need to return to equation (\ref{eq01}) for the energy loss.
Perturbative QCD predicts that the radiative energy loss depends 
quadratically on the medium thickness $L$ \cite{BDMPS}. 
As pointed out by Baier et al. \cite{BDMS}, this holds for the average
energy loss $\overline{\Delta E}$ of a given parton, but the average
energy loss of observed partons with fixed transverse momentum $p_T$
has a different scaling. This is so, because the energy loss distribution
$D(\epsilon)$ is strongly skewed toward small values of $\epsilon$ by the
steeply falling $p_T$-spectrum of fast partons. In fact, the average
shift of the spectrum due to the energy loss, here called the effective
energy loss, is given by \cite{BDMS}
\begin{equation}
\Delta p_T \approx \alpha_s L \sqrt{\pi {\hat q} p_T / \nu} \, ,
\label{eq10}
\end{equation}
where ${\hat q}$ encodes the ``scattering power'' of the medium, which
is proportional to the density. For an expanding medium, the expression 
${\hat q}L^2$ must be replaced with
\begin{equation}
{\hat q}_0 L_{\rm eff}^2 = 2 {\hat q}_0 \int_0^L \tau d\tau\, 
           \frac{\rho({\bf r}(\tau),\tau)}{\rho({\bf r},0)} \, ,
\label{eq11}
\end{equation}
where ${\bf r}(\tau) = {\bf r} + {\bf v}\tau$ denotes the position of 
the fast parton in the medium at time $\tau$, and ${\hat q}_0$ is a
function of the transverse position ${\bf r}$ at which the jet is
produced. We thus can make contact with (\ref{eq01}) by writing
\begin{equation}
\Delta p_T =  \eta' L_{\rm eff} \sqrt{\rho({\bf r},0) p_T / \nu} 
\label{eq12}
\end{equation}
with the constant $\eta' = \alpha_s \sqrt{\pi{\hat q}_0/\rho(0)}$,
which does not depend on ${\bf r}$.

We will denote the scaling law (\ref{eq12}) for the energy loss as BDMS.
In our following numerical study we have explored two other scaling laws.
The first one is the Bethe-Heitler (BH) scaling law \cite{Jeon02,Mus98}
\begin{equation}
\Delta p_T = \eta p_T \int_0^L d\tau\, \rho({\bf r}(\tau),\tau)
             \equiv \eta p_T (L\rho)_{\rm eff} 
\label{eq13}
\end{equation}
corresponding to $\mu = 1$.  The second scaling law is
\begin{equation}
\Delta p_T = \eta p_T \sqrt{(L\rho)_{\rm eff}} \, ,
\label{eq14}
\end{equation}
which we will call the (RW) scaling law. It could be interpreted 
as describing a random walk in $p_T$ as the fast parton traverses the
medium, with some interactions resulting in an energy gain and others
in a loss of energy. 

\begin{figure}
\resizebox{0.65\linewidth}{!}{\rotatebox{-90}
          {\includegraphics{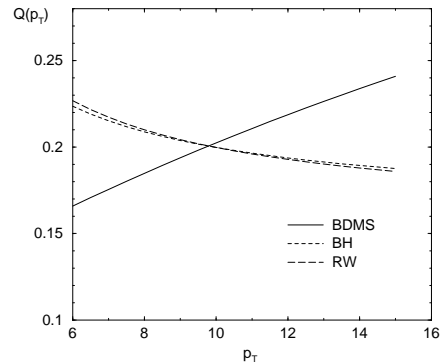}}}
\caption{Dependence of the quenching factor $Q$ on $p_T$ for central
         collisions.  The parameter $\eta$ is chosen such that
         $Q(p_T)\approx 0.2$ for $p_T = 10$ GeV/c in each case.
         The scaling laws (BH, RW) exhibit stronger quenching with
         increasing $p_T$, in agreement with preliminary RHIC data, 
         in contrast to the BDMS law. Equation (\ref{eq05}) provides
         a good description of the dependence on $p_T$ seen here.}
\label{fig2}
\end{figure}

We begin the discussion of our numerical results for the quenching factor 
$Q$ with its dependence on the  transverse momentum of the fast parton,
shown in Fig.~\ref{fig2}. The QCD-motivated BDMS law (solid line) and 
the other two scaling laws exhibit clearly different behaviors. This 
reflects the different $p_T$ scaling of the energy loss in these models 
(linear for BH and RW; square-root for BDMS). The data from the RHIC 
experiments \cite{PHEN02,STAR02,PHEN-QM1} suggest that the quenching 
first becomes stronger with increasing momentum, reaches a minimum, and
finally begins to diminish. This would indicate that the BDMS law only 
applies at high $p_T$, and that other laws govern the energy loss at
lower $p_T$ \cite{Jeon02} or hadron production is not dominated by parton
fragmentation in this kinematic region. We note that the dependence of $Q$ 
on $p_T$ is well described by the analytical formula (\ref{eq05}).

\begin{figure}
\resizebox{0.65\linewidth}{!}{\rotatebox{-90}
          {\includegraphics{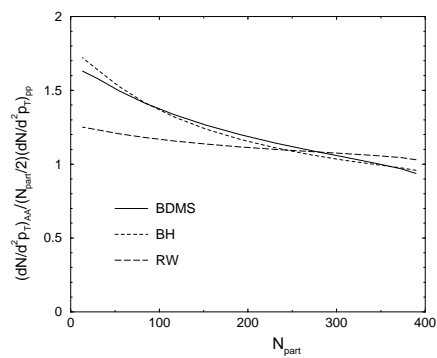}}}
\caption{Quenched hard parton yield divided by half the number of 
         participant nucleons as a function of $N_{\rm part}$
         for $p_T = 10$ GeV/c. The values of the stopping power
         strength parameters are $\eta=0.06$ (RW), $\eta=0.017$ (BH),
         and $\eta'=1.1$ (BDMS).}
\label{fig3}
\end{figure}

The impact parameter dependence of the quenching factor is shown
in Fig.~\ref{fig3}, plotted as the yield per half the number of 
particpant nucleons against the particpant number $N_{\rm part}$. 
The unquenched jet yield, scaling with the number of binary NN 
collisions, would increase relative to $N_{\rm part}$. As the 
figure shows, the quenching counteracts this increase, and the
yield per participant actually falls for the BDMS and the BH laws
as the collision centrality increases. An approximately flat behavior, 
as observed in the PHOBOS experiment \cite{PHOB-QM}, is only found 
for the RW scaling law.

\begin{figure}
\resizebox{0.65\linewidth}{!}{\rotatebox{-90}
          {\includegraphics{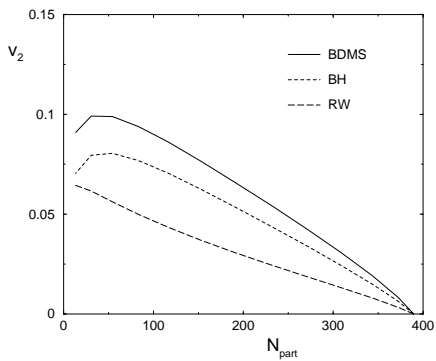}}}
\caption{Elliptic flow parameter $v_2$ as a function of the number
         of participants for the three energy loss models. The
         values of the parameters as the same as in Fig.~\ref{fig3}.}
\label{fig4}
\end{figure}

For noncentral collisions, the quenching factor $Q$ is a function
of the azimuthal emission angle, because the geometry is not
axially symmetric. This is known to lead to an angular asymmetry
of a quadrupole shape in the spectra of high-$p_T$ particles
\cite{Wan01,Gyu01}. The elliptic flow parameter $v_2$ is defined 
as the Fourier component proportional to $\cos(2\phi)$ of the
angular distribution of particles with respect to the scattering
plane \cite{Pos98}. We find (see Fig.~\ref{fig4}) that the values 
of $v_2\le 0.1$ obtained for all three scaling laws are significantly 
smaller than the measured values ($v_2\approx 0.2$) for semicentral 
and peripheral collisions \cite{STAR02a}. However, the calculated 
$v_2$ for partons would be large enough to explain the measured 
elliptic flow of hadrons, if the hadrons were produced by 
recombination in the $p_T$-range of the RHIC data \cite{Molnar}.

\begin{figure}
\resizebox{0.65\linewidth}{!}{\rotatebox{-90}
          {\includegraphics{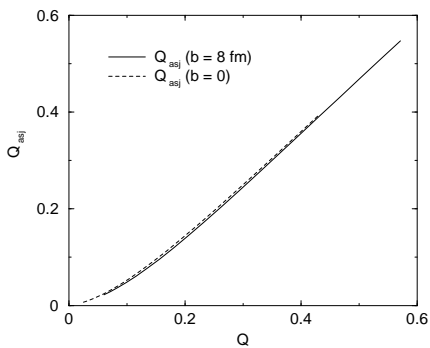}}}
\caption{Incremental away-side jet suppression factor $Q_{\rm asj}$
         as function of the ``same-side'' jet suppression factor
         $Q$ for the BDMS energy loss law.}
\label{fig5}
\end{figure}

As observed hadrons from hard partons preferentially originate
from the surface region facing the detector, the parton emitted
in the opposite direction has to traverse more material and thus
endures an even larger energy loss. This leads to an additional 
suppression of the away-side jet and its leading hadron spectrum. 
The dependence of the incremental away-side hadron suppression
factor $Q_{\rm asj}$ on the primary suppression factor $Q$ is 
shown in Fig.~\ref{fig5} for the BDMS scaling law for two impact
parameters ($b=0$ and $b=8$ fm). There appears to be a universal 
relationship, which is linear for $Q\ge 0.2$.

Several conclusions can be drawn from our results. First, the
momentum dependence of the BDMS energy loss formula does not
seem to agree with some of the RHIC data. A linear dependence 
of the average energy loss on $p_T$ is in better agreement with 
the data in the $p_T$ region explored so far \cite{Jeon02}. 
Also, the dependence of the effective
momentum loss $\Delta p_T$ on the medium thickness $L_{\rm eff}$ 
predicted by BDMS does not yield a scaling of the charged 
hadron yield with participant number as seen in the PHOBOS data 
up to $p_T = 4.25$ GeV/c; only the RW scaling law yields such a
dependence. This may indicate that energy gain and loss mechanisms 
are competing in this momentum range \cite{WW01}. 
Another explanation may be that hadrons at intermediate 
momenta are not produced by fragmentation of fast partons, but by 
other processes, such as parton recombination. The observed magnitude
of the elliptic flow lends support to this interpretation.
Finally, we have found a universal relation between the same-side
and away-side suppression factors for the BDMS law, which can be
tested experimentally.

{\em Acknowledgments:} This work was supported in part by a grant
from the U.~S.~Department of Energy (DE-FG02-96ER90945). I thank
S.~A.~Bass, X.-N.~Wang, and K.~Rajagopal for valuable comments on 
the manuscript.

\end{document}